\documentclass{article}

\usepackage{arxiv}

\usepackage[utf8]{inputenc} 
\usepackage[T1]{fontenc}    
\usepackage{hyperref}       
\usepackage{url}            
\usepackage{booktabs}       
\usepackage{amsfonts}       
\usepackage{nicefrac}       
\usepackage{microtype}      
\usepackage{lipsum}		
\usepackage{graphicx}
\usepackage{doi}

\usepackage[figuresright]{rotating}
\usepackage{pdflscape}
\usepackage{longtable}

\title{An improved tile-based scalable distributed management model of massive high-resolution satellite images}

\author{ {\hspace{1mm}Yosra Hajjaji}\\
	RIADI Laboratory\\ 
	National School of Computer Science\\
	University of Manouba\\
	Manouba 2010, Tunisia \\
	\texttt{hajjajiyosra05@gmail.com} \\
\And
	{\hspace{1mm} Wadii Boulila}	 \\
	RIADI Laboratory\\ 
	National School of Computer Science\\
	University of Manouba\\
	Manouba 2010, Tunisia \\
	\texttt{wadii.boulila@riadi.rnu.tn} \\

\AND
   {\hspace{1mm} Imed Riadh Farah}  \\
	RIADI Laboratory\\ 
	National School of Computer Science\\
	University of Manouba\\
	Manouba 2010, Tunisia \\
	\texttt{imedriadh.farah@isamm.uma.tn} \\
}

\begin{document}
\maketitle

\begin{abstract}
The amount of remote sensing (RS) data has increased at an unexpected scale, due to the rapid progress of earth-observation and the growth of satellite RS and sensor technologies. Traditional relational databases attend their limit to meet the needs of high-resolution and large-scale RS Big Data management.
As a result, massive RS data management is currently one of the most imperative topics. To address this problem, this paper describes a distributed architecture for big RS data storage based on a unified metadata file, pyramid model, and Hilbert curve for data composition and indexing using NoSQL databases (i.e, Apache Hbase). In this paper, a Hadoop-based framework in AzureInsight cloud platform is designed to manage massive RS data in a parallel and distributed way. Experimental results prove that our method has the potential to overcome the weakness of traditional methods. The proposed model is suitable for massive high-resolution image data management.
\end{abstract}


\section{Introduction}
Remote sensing (RS) has been extensively applied in several fields, such as disaster assessment, climate, agriculture, environmental monitoring, and many other applications \cite{Atli2012very}. Over the past few years, due to the rapid improvement in computer technologies and sensors, RS data have proliferated over the world.\\
An important sequence of sensors with high-resolution (spectral, spatial, and temporal) has been launched, which led to an exponential rise in RS data, making available a huge volume of spatio-temporal information, \cite{He2017spatio, Boulila2009improving}. A tremendous volume of RS data is becoming increasingly abundant from many satellite data centers such as NASA Open Government Initiative in addition to other geo-spatial data sources. Currently, over 1000 RS sensors have been initiated \cite{Zhu2018a} and the volume of data accumulated by one satellite data-center is piling up at an exponential rate (terabytes daily).

The volume of RS image data becomes massive and continues to increase exponentially. For example, the level of EOSDIS (Earth observation data of the European Space Agency) and the ESDIS (Earth Science Data and Information System) had exceeded the 1.5 PB and 7.5 PB, respectively \cite{Gamba2011Foreword,Chi2016Big}. As a result, big data has become a hot topic in the RS field and it has been commonly known, today, as "remote sensing big data" \cite{Gamba2011Foreword}.
Such a large volume of data has led to serious difficulties in data storage and processing, which cause weaknesses in current data management systems and data organization methods in order to meet the new application requirements. According to both academic and business fields, we found that the main challenges are how to efficiently store, manage and quickly retrieve such massive data \cite{Chi2016Big,Masood2020a}. These issues have become increasingly important owing to the concern of spatial information for researchers, business applications, and agencies \cite{Hajjaji2018performance}. Moreover, over recent years, many researchers, scientists, and engineers asked for new methods for fast RS image data management (i.e., real-time or nearly-real time) \cite{Hajjaji2021Big,Boulila2021RS}. \\ 

Overall, databases are amongst the best manners to store data. However, since the quantity of spatial data is increasing rapidly, the traditional relational database could no longer deal with massive data issues. They either cannot handle such data volume of data or confronting performance difficulties, particularly the problem of data storage and efficient data access when the volume (data scale) and the velocity (speed up at which data is generated) of data raises at a remarkable rate. Faced with the two previously mentioned challenges, it becomes appropriate to adopt new technologies.
For the sake of the problem related to efficiency manage RS big image data, a distributed model for massive high-resolution image data storage is designed.
First, a unified metadata file model is proposed to transform metadata into a uniform format before data importation across several distributed data centers. Then, two classification models are used; the pyramid model and the Hilbert-SFC-based data indexing mechanism for optimal query and access of RS images. Additionally, a NoSQL database management system (Hbase) is used to store the data \cite{b14}. This database is distributed, so the stored data are distributed on different data nodes. Moreover, we used the Hadoop framework for RS big data management and processing \cite{Sadalage2016nosql}.
\section{Related works}
In this section, we present an overview of existing researches studies that answer the question of how efficiently manage massive RS image data. We mainly put the light on their architectures, subdivision models, index structures, and used tools and methods. We also checked both spatial correlation and metadata standardization aspects in order to build the proposed method (see Table \ref{table.1}). 

\begin{sidewaystable}
\begin{landscape}
\begin{longtable}{|p{0.5cm} |p{3cm} |p{3cm} |p{2.80cm} |p{4.75cm} |p{2cm}|p{3cm}|}
\caption{An overview of existing primary studies.} 
\label{table.1}
\\
\hline\endhead  
\hline\endfoot

\textbf{Ref.} & \textbf{Physical storage architecture}&  \textbf{Subdivision model} & \textbf{Spatial correlation} & \textbf{Index structure} & \textbf{Metadata standardization} & \textbf{Tools \& methods} \\\hline

\cite{Lie2013massive} & Distributed storage & Pyramid map & Yes & Hbase key-value  &  No &  Hadoop, Hbase, Mapreduce   \\ \hline

\cite{Li2016Integration} & Local server &  GeoSOT ($2^n$-tree) & Yes & Inverted index  & Yes  & Kingbase enterprise server database system Oracle platform   \\\hline

\cite{Yang2017Efficient} & Single and distributed storage  & Pyramid map /
TMS (Tile Map Service) & Yes & Hbase key-value  &  No & Hadoop, HDFS, Hbase, Spark RDD   \\ \hline
\cite{Jing2018an} & Distributed storage  & Image pyramid + uniform grid division  &  Yes &  The grid index and the Hilbert curve &  No &  Hbase, Mapreduce, MRHbase, MySQL \\ \hline
\cite{Hajjaji2018performance} & Distributed storage & Pyramid map & Yes & Hbase key-value &  No & Hbase, Cassandra, MongoDB \\
\hline
\cite{Wang2019a} & Distributed cluster architecture  & GridFS mechanism  &---  & MD5 & No  & MongoDB, PostgreSQL, WiredTiger engine  \\\hline
\cite{Xu2020Science} & Distributed cluster architecture & Quad Tree & Yes & Hbase key-value based on quad tree + Hilbert curve & Yes  & Hadoop, Hbase, HDFS,
Spark, OpenStack cloud \\\hline
\cite{Wang2020a} & Distributed cluster architecture &  Google S2 & Yes & Hbase key-value based on Hilbert curve & No  & Hadoop, Zookeeper, Hbase  \\\hline
\cite{Yang2020a} & Distributed storage &  Google S2 & Yes & Hbase key-value + Kylin  &  Yes &  Hbase, Zookeeper, Hive, Kylin  \\ \hline
\end{longtable}
\end{landscape}
\parbox{20cm}{---: Not mentioned}
\end{sidewaystable}

In 2013, Yuehu Liu et al. \cite{Lie2013massive} intended to efficiently manage massive RS data with a non-traditional method since traditional methods show their limit of being expensive and hard to extend. Therefore, in order to meet the requirements of a parallel and scalable processing model, the authors designed a new method based on pyramid map, Hbase, and MapReduce for data storage data processing respectively. Results show that data importing and processing speed increases obviously along with the cluster of Hbase. However, it has been indicated that more tests with the best experimental environment are required. In addition, a good knowledge of Hadoop parameters is required as it can affect the performance results.

In 2016, Shuang Li et al. 
\cite{Li2016Integration} provided a method of RS data integration and management based on the Geosot subdivision model. The authors confirmed that their approach offers a more effective storage management program for existing storage centers and management systems, a low cost of creating a logical subdivision index, and a query speed improvement for RS data. This approach needs to be tested in a distributed manner and to efficiently improve the accuracy. In 2017, Mengzhao Yang et al. \cite{Yang2017Efficient} have implemented a method for massive RS image using Spark-based pyramid model. According to experiments and results, this study shows that the efficiency of massive image data storage can be obtained thanks to an optimization of the image tile parameter. Also, the authors proved that their method offers a better throughput rate and construction performance compared with MapReduce and Hadoop. However, many interesting directions require to be explored, such as applying some practical applications like image retrieval, image detection, and image classification based on this model.
In 2018, Jing and Dongxue \cite{Jing2018an} and Yosra et al. \cite{Hajjaji2018performance} designed a storage model of RS image data based on open source big data technology and pyramid map for image data subdivision aiming to improve the issues of low efficiency and scalability deficit of traditional processing models. The outcomes of their methods show that it can strengthen both data writing, query speed, and high scalability. Besides, the improved index used in \cite{Jing2018an} has a high spatial proximity in the Hbase table.
In 2019, Shuang Wang et al. \cite{Wang2019a} presented a distributed storage and access method for big RS data based on GridFS mechanism to improve the I/O speed, horizontal expansion and to enhance the low query performance in the process of massive RS data management. This approach assured a good scalability and a random access performance thanks to the Sharding technology. However this contribution need to study the influence of the number of data nodes on the performance of the distributed system. \\
Recently, in 2020, Yinyi Cheng et al. \cite{Xu2020Science} proposed ScienceEarth, a framework for large-scale RS data storage, management, and processing in cluster-computing in a cloud environment. According to their test results, the availability and computation of RS data prove that their approach can efficiently retrieve and process RS data. However, this efficiency can be affected by the workload situation in the cloud. ScienceGeoIndex needs to be optimized to provide standard map services. The platform needs to support vector data and to include machine learning (ML) and deep learning (DL) frameworks under consideration. In the same year, Wang et al. \cite{Wang2020a} and Yang et al. \cite{Yang2020a} suggested a method for RS image data storage based on Google S2 and Hbase. The authors propose to greatly improve tile storage and query efficiency, and resolve the issues of data organization and data sharing due to the use of the various format of similar spatial object respectively. Nevertheless, these contributions need to be applied to other distributed databases, data retrieval, data analysis, focus on the lost tile reconstruction, and mining fine-grained RS data. 

\section{Background}
\subsection{NoSQL and Hbase}
Although traditional approaches (i.e., Relational Database Management System (RDMS)) fill an important position in the data management area, massive RS image data bring out a new concern about data management and processing \cite{Xiaofeng2013big, Boulila2017Sensitivity}. 
Due to the ACID (atomicity, consistency, isolation, and durability) properties and users' demand for optimal solutions, RDMSs reach their bottlenecks in massive data management. NoSQL databases have emerged as a solution to the above problems \cite{Corebellini2017Persisting,Shen2013Survey}. 
This new technology proved to be efficient in massive data storage and management, due to its feasible data models, high I/O performance, and powerful scalability for a massive volume of data.  Hbase is a freely accessible distributed database that runs on the top of the Hadoop Distributed File System (HDFS) \cite{Hajjaji2018performance}. It enables a fault-tolerant manner of storing data, random, consistent, and real-time access to big data. Hbase model stores data in tables based on rows and columns. A unit of data comprises several rows and specific columns in the table named as an element, where every element retains several variations of the similar data presented by different timestamps \cite{Xu2020Science, Wang2020a}. In the Hbase table, a column family is composed of a set of columns logically related, and it should be determined before using the table. Then, the columns are generated dynamically throughout utilization. In Hbase, data are stored through a dictionary according to a row keys order. Each row key is unique, it is the only identifications that define the table files. It also plays the role of the primary key for data retrieval.

\subsection{ The technology of multi-resolution pyramid }

\begin{figure}[htbp]
\centerline{\includegraphics[width=8cm, height=4cm]{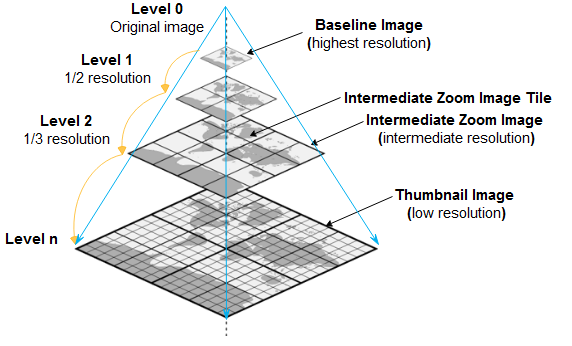}}
\caption{ Four levels RS image tiled pyramid.}
\label{Fig.1}
\end{figure}

The multi-resolution pyramid model is an appropriate way for both large-scale and high-resolution RS image data management and organization. The structure of this model is simple to handle and working as a kernel data structure over plenty of virtual globe systems (i.e., Google Earth). The design of the pyramid image is a block-oriented image decomposition based on a hierarchical structure that represents a big-scale RS image at several Level Of Details (LOD) according to the image resolution starting from the highest to the lower level (check Figure~\ref{Fig.1}). 

Figure~ \ref{Fig.1} presents an example of a four-level pyramid. The pyramid level means the count of increased image resolution levels generated during the pyramid construction. The more high-resolution the image is the more levels of the pyramid we get. The bottom of the pyramid is the original image. If we consider an image with a resolution of $1024\times 1024$ at the bottom of the pyramid. By resampling the initial image, the resolutions rapidly decreased to $512\times 512$, $256\times 256$, and $128\times 128$ respectively with the growth of the pyramid layers.

\subsection{Standardization of RS metadata archiving model} 

RS metadata plays a crucial role in complementary studies of earth observation. Efficient management of this metatada can facilitate the application and exchange of RS knowledge \cite{Yang2017Efficient}. In this study, RS metadata is the descriptive information of the satellite image data. This metadata is created to save attribute information. Nevertheless, the metadata file's components are diverse, which brings complications to RS metadata unified management. In the ZY-3 metadata file, for example, the Satellite-ID and Sensor-ID are the corresponding fields. While in the Landsat8 metadata file, the "SENSOR-ID" and the "SPACECRAFT-ID" fields are the sensor identifier in the satellite and the satellite identifier of the image respectively. Therefore, building up a unified archiving model for RS data management is imperative.\\
In this study, we will use the ISO 19115-2 2009 a standard of Geo-information metadata. It is the second section of the ISO 19115 and an expansion of the image and the grid data. It has been merged into the global metadata warehouse CMR (Common Metadata Repository) \cite{ Huang2018rapid, Molina2013Earth}, and it became a standard for data share, data integration, and data recovery between Geo-data centers and international geographic information organizations. 
In this paper, we will use a uniform standard format model for archiving RS metadata and determine its fields as depicted in Table \ref{table:2}. This standard is founded on the survey and investigation of existing metadata standards including ISO (International Organization for Standardization) 19115 geographical information metadata standard and CSDGM (Content Standard for Digital Geospatial Metadata) ISO 19115:2003 cor.1 2003, in addition to the metadata structure of various RS data resources.

\begin{table}
\caption{Standard archival model for RS metadata}
\begin{center}
\begin{tabular}{|p{2cm}|p{4cm}|p{5cm}|}
\hline
\textbf{Classification}	& \textbf{Field name}	& \textbf{Description} \\
\hline
MetadataInfo & Creation & Time of metadata creation\\
& Last update & Time of last update \\
\hline
ImageInfo & ImageName & Image data name\\
& Tilescodes & \\
& Level & \\
& Satallite-ID & Satellite identifier\\
& Sensor-ID & Sensor identifier\\
& ReceiveTime & Time of data reception\\
& Satallite-ID & Satellite identifier\\
& ReferencesysitemID & Reference coordinate system\\
& Time-beginposition  & Start time\\
& Time-endposition & Stop time\\
& ProductLevel & Product level\\
& ProductFormat & Product format\\
& Time-endposition & Stop time\\
& SpatialResolution & Spatial resolution\\
& Time-endposition & Stop time\\
& ProcessingLevel &Pprocessing level\\
& Centerlongtude& Center point longitude\\
& Centerlatitude & Center point latitude\\
& TopLeftLatitude & Latitude in the upper-left corner \\
& TopLeftLongitude  &  Longitude in the upper-left corner\\
& TopRightLatitude  &  latitude in the upper-right corner\\
& TopRightLongitude  &  Longitude in the upper-right corner\\
& BottomRightLatitude  &  Latitude in the lower-right corner\\
& BottomRightLongitude  &  Longitude in the lower-right corner\\
& BottomLeftLatitude  &  Latitude in the lower-left corner\\
& BottomLeftLongitude  &  Longitude in the lower-left corner\\
& FilePath & Path\\
& SceneRow & Row\\
& CloudPercent & Cloud cover volume\\
& DataLink & Data download URL\\
& DataProvider & Data provider\\
& DataOwner & Data owner\\
\hline
\end{tabular}
\label{table:2}
\end{center}
\end{table}

\section{The proposed system}
To ensure the distributed storage of high-resolution satellite image data, we designed an Hbase model based on tile pyramid image data structure. By indexing the metadata describing the satellite image data, the storage and access have been efficiently improved.

\subsection{Pyramid model for image division}

In general, the size of satellite images is more than 300 MB. After dividing the image, the volume of data is expanded by 1/3. Therefore, the size of mosaic tiles data can exceed 10 GB or 100 GB \cite{ Chen2014Pyramid, Li2019a}. The pyramid model based on image block technology helps to efficiently store and access RS data \cite{Wan2009Research}.

The most common method to organize image data blocks is to build up a hierarchic and multiscale pyramid model with a global sub-division grid image. Therefore, end-users could easily reach the image blocks of particular spatial regions and specific resolution levels as needed \cite{Cheng2020big}.\\
Generally, most RS images are composed of multiple bands. In this paper, we are dealing with high-resolution satellite images (e.g., multi-spectral images), where each image has several bands. Every single band is a blend of several pyramid layers, where a range of blocks may be presented by a pyramid layer. 

\begin{figure}[htb]
\centerline{\includegraphics[width=8cm, height=7cm]{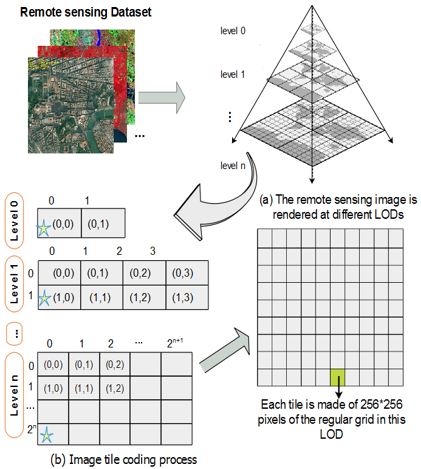}}
\caption{Pyramid model of bloc image.}
\label{Fig.2}
\end{figure}

First, the original image is at the bottom of the pyramid, which is denoted as "level 0" (highest level of data), then each layer is split into equal chunks. The image pyramid is illustrated in Figure \ref{Fig.2} (a). Any block is named a tile. A tile is constituted by level, row, and col. According to these adjustments, this is how tile data encoded. The encoding process is demonstrated in Figure \ref{Fig.2} (b).

The partitioning image method is as follows; initially, based on the multi-resolution pyramid model, the high-resolution satellite image must be divided into disparate zoom layers (see Figure~ \ref{Fig.1}). Then, as stated in the theory of partition, each image of the pyramid model will be divided into multiple layers and rectangle tiles that all have the same pixel dimensions. For example, the tile size is $128\times$128 in ArcSDE (SDE for Spatial Database Engine) and the tile size is $256\times 256$ in Oracle GeoRaster \cite{Xie2007Georaster}. By default, the size of the block in the proposed system is $256\times 256$ \cite{Lu2011Review}. Second, in the pyramid layer, a block can be identified by a row number and a column number ($rownu$, $colnu$) respectively. Corresponding values at pyramid level $\kappa$, can be calculated using the following equations, specifically, with a given longitude $\lambda$ and latitude $\varphi$:

\begin{equation}
rownu= \left[ ((\varphi + 90)/ (180/2^k\right] \bmod 2^k
\end{equation}
\begin{equation}
colnu= [((\varphi + 180)/ (180/2^k)] \bmod 2^{k+1}
\end{equation}

Moreover, the geographic area of a tile could be intended through the ($rownu$, $colnu$) numbers and the pyramid level $k$ as follows: 

\begin{equation}
west= ((rownu \bmod 2^{k+1}) \times 180/2^k) -180
\end{equation}
\begin{equation}
east = west + 180/2^k
\end{equation}
\begin{equation}
south = ((colnu \bmod 2^k ) \times 180/2^k) - 90
\end{equation}
\begin{equation}
north= south + 180/2^k
\end{equation}

\begin{figure*}[htbp]
\begin{center}
\includegraphics[width=0.7\textwidth]{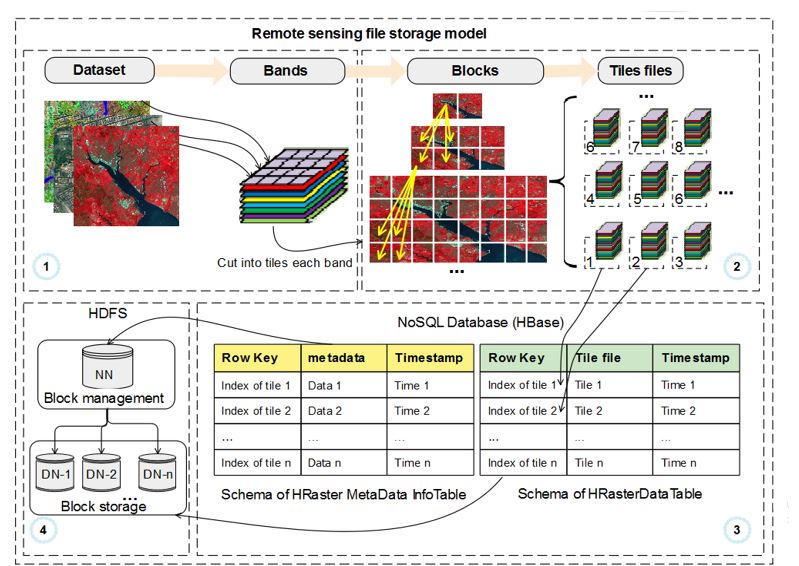}
\end{center}
\caption{RS data system with a HDFS and Hbase.}
\label{Fig.3}
\end{figure*} 

\subsection{Image data storage model} 
In order to efficiently manage high-resolution RS images, many researchers used an image structure for data division (i,e., \cite{Yang2017Efficient}). Based on the previous research works, we design a novel model based on Hbase and HDFS for image data storage and access, taking advantage of distributed database benefits. The overview of the proposed model for image data storage is shown in Figure \ref{Fig.3}. The entry of the suggested system is a set of RS image data. In our context, we are dealing with high-resolution (spatial, spectral, and temporal) images (e,g., multi-spectral images, hyper-spectral images, high-spatial-resolution images, and high-temporal resolution images). Each one of these images has several bands as depicted in Figure \ref{Fig.3} (1)). We divide each band into different layers and blocks according to the process of building the pyramid model (see the first section). 
Therefore, a set of pyramids is established for each band.\\
In order to decrease the data transmission across the worker nodes, we used the band sequential format (BSQ) to keep all data of the same location, from all bands, in the same block (see Figure \ref{Fig.3} (2)). Then, we extract those blocks for each pyramid and store them in the Hbase database. We designed a tile-based and scalable high-resolution data storage system based on Hbase and HDFS. Two tables, called HRaster MetaDataInfoTable and HRasterDataTable, were designed to store the metadata and the RS images data blocks, respectively. To reduce the query time while data retrieval, the metadata column in the HRaster MetaDataInfoTable table is used to store the metadata of each layer of the image. Every time a novel RS image is imported, a new file of metadata will be written in HRaster MetaDataInfoTable and a new RasterDataTable table will be generated to save the tile file chunks (see Figure \ref{Fig.3} (3)).\\ 
We create an Hbase database on the top of the HDFS so that it will leverage both load balance and security provided by HDFS. Hbase ensures storing massive volumes of data, in addition to efficient data inquiring based on the row keys. The master/slave architecture of HDFS is presented in Figure \ref{Fig.3} (4). The NameNode manages the file system and the metadata of the overall RS data, while DataNodes stores the block data.
\begin{figure}[htbp]
\begin{center}
\includegraphics[width=7cm, height=6cm]{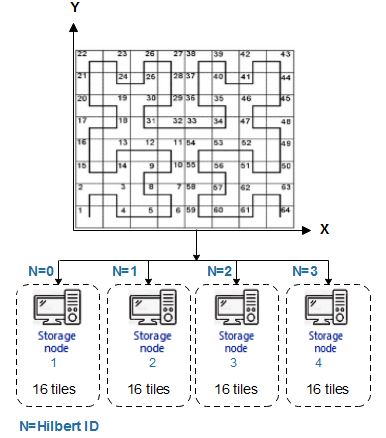}
\end{center}
\caption{Tiles distribution using BPS.}
\label{Fig.4}
\end{figure} 

To improve the efficiency of the distributed database, we used the balanced placement strategy (BPS) and the periodic storage strategy (PSS) proposed in \cite{Wang2020a}. First, according to the BPS strategy, the tiles will be stored uniformly during their distribution among the nodes (see Figure \ref{Fig.4}). This strategy helps to avoid data skew and hot-spotting and improves query efficiency since the tile will be stored together according to their spatial correlation. Second, the PSS strategy aggregates the tile according to the period to reduce data blocks while storing tiles of the same image. Therefore, the proposed strategy improves the time interaction within data blocks in addition to the performance of data management and access.

\subsection{RS tile data index}
There are many available definitions of the space-filling curve (SFC) based on different mapping options (i.e., Morton, Peano, and Gray) \cite{Sagan2012Space} \cite{Borrell2018parallel}. Compared with other linearization curves, Hilbert SFC seems to be excellent and more stable \cite{Borrell2018parallel}. Therefore, we propose to apply the Hilbert SFC method to fill the grid of each image and to index each tile. This method ensures excellent spatial proximity of the data; it maps two-dimensional spatial positions into one-dimensional space and gives cells unique codes \cite{Wang2020a}. 
Therefore, tiles with corresponding Hilbert codes are saved with each other.
\begin{figure}[htbp]
\centerline{\includegraphics[width=8cm]{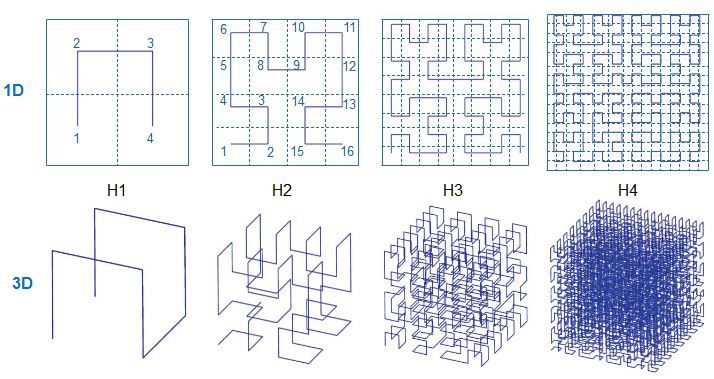}}
\caption{The filling process of Hilbert curve for tiles data index.}
\label{Fig.5}
\end{figure}
Figure \ref{Fig.5} illustrates the generation process based on step-wise geometric recursion, of four order Hilbert curve (H1,  H2, H3, and H4) for both 1D and 3D data. Following the pyramid model design and Hilbert curve indexation model, we define the principle of tile indexation and row keys of every tile in the database.
\begin{figure}[htbp]
\centerline{\includegraphics[width=9cm]{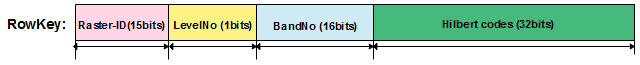}}
\caption{Hbase key value structure}
\label{Fig.6}
\end{figure}

We used a sample of a row-key model for tile data constituted four principal parts (see Figure ~\ref{Fig.6}). Bits from 0 to 15 are used to store RasterID, bits from 15 to 16 are used for pyramid level number, bits from 16 to 32 are used for image band number and the last part of the row-key is 32 bits used for Hilbert code, which points out the tile spatial location inside the image. Through RS data indexation, we can easily perform specific queries. Moreover, based on the RS tile data index, we can easily access RS data stored in the database for data processing and analysis.

\section{Experimentation and results}
\subsection{Dataset}
The experimental dataset includes high-resolution images and metadata. The considered dataset is composed of free satellite images coming from different data centers, which includes three images from Spot-5 and Spot-6/7, three images from Quickbird, and four images from GeoEye. All the images are in blue, green, red (RGB), and near-infrared at different spatial resolutions. Table \ref{table:3} presents the description of each data product. RS images are organized into three groups as a test pattern,  separated from small to big size groups, from 1GB, 4GB, and 8GB respectively.

\begin{table}[h]
\caption{RS images data details}
\begin{tabular*}{\hsize}{@{\extracolsep{\fill}}lllll@{}}
\toprule
\textbf{File groups}&\textbf{Quantity of images} &\textbf{Satellite }& \textbf{Number of bands} & \textbf{Resolution}  \\
\hline
Group 1 & 3 images &Spot-5, Spot-6, Spot-7 &  4 bands & 6m-10m \\
\hline
Group 2 & 3 images &QuickBird  &  4 bands& 2.4m \\
\hline
Group 3 & 4 images &GeoEye  &  4 bands& 1.65m \\
\hline
\end{tabular*}
\label{table:3}
\end{table}

\subsection{Experimental Environment}
In our experiments, virtual machines are deployed
in Azure HDInsight cloud platform. Azure HDInsight is an open-source, full-spectrum that offers analytics management service in the cloud. It runs the most popular open-source frameworks like Apache Hadoop, Apache Hive, Apache Kafka, Apache Spark, Apache Storm, and others \cite{b36}. The node features details of the cluster are depicted in table \ref{table:4}. Each instance machine is equipped with Hadoop version 2.7.3, ZooKeeper 3.4.14 version, and Hbase version 1.4.13.

\subsubsection{Experiment A:  Parameters optimization of image tile:}
In this experiment, two different parameters, the tile size, and the data size were selected to compare the slice time per response time. We considered two storage models at several scales and time storage as measurements for the performance comparison.

\begin{table}[h]
\caption{HDInsight Azure VMs Node description}
\begin{tabular*}{\hsize}{@{\extracolsep{\fill}}lll@{}}
\toprule
Node type & Head Node & Worker node \\
\hline
Instance & A8 v2 & A4 v2  \\
\hline
Node size & 8 cores, 16GB RAM  &  4 cores, 8GB RAM  \\
\hline
OS & \$0.40/hour &  \$0.191/hour    \\
\hline
HDInsight Price & 	\$0.068/hour & \$0.136/hour       \\
\hline
Total Price  & 	\$0.536/hour  &  \$0.259/hour   \\
\hline
\end{tabular*}
\label{table:4}
\end{table}

\begin{figure} [htbp]
\centerline{\includegraphics[width=13cm]{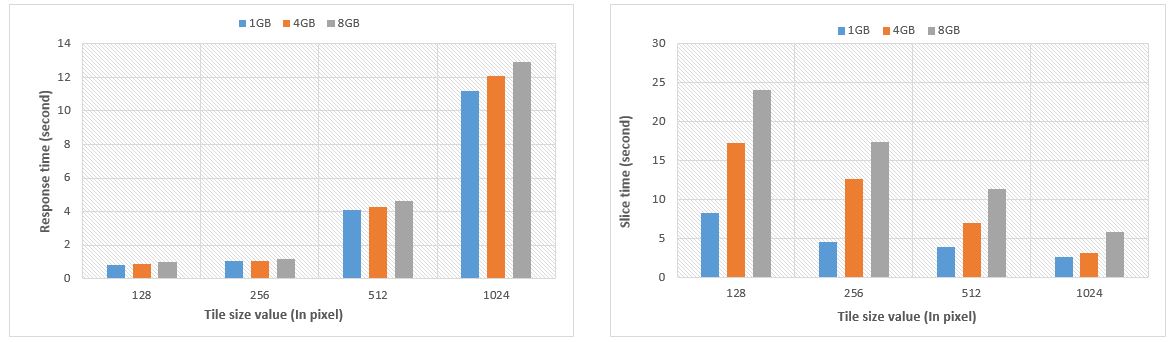}}
\caption{Response time and slice time of 4 construction methods with different tile size value and data size respectively}
\label{Fig.7}
\end{figure}

From Figure \ref{Fig.7}, we notice that once the tile size is getting bigger, the response time is getting longer and vice versa. In addition, we also notice that as much as the RS data size increases, the division time is increasing too. As a result, if we take into account the fast query response time requirement, we need to carefully choose the value of the tile size parameter. Furthermore, when the RS image data size is large, the efficacy of image processing can be improved. In this case, an equilibrium parameter value needs to be selected to meet both users' demands of rapid data access and image processing.

\subsubsection{Experiment B: Scalability on cluster and data size:}

In this experiment, we aim to analyse the efficiency of the suggested distributed management model for massive high-resolution images.

First, to measure the effect of the number of data node numbers in the cluster at the data writing time, the ingestion performance of the database is tested. The traditional oracle spatial database is used with the proposed Hbase model for comparative analysis. Therefore, the total size of data is held constant while the number of data nodes increased from 1 to 12 nodes.\\

\begin{figure} [htbp]
\centerline{\includegraphics[width=13cm]{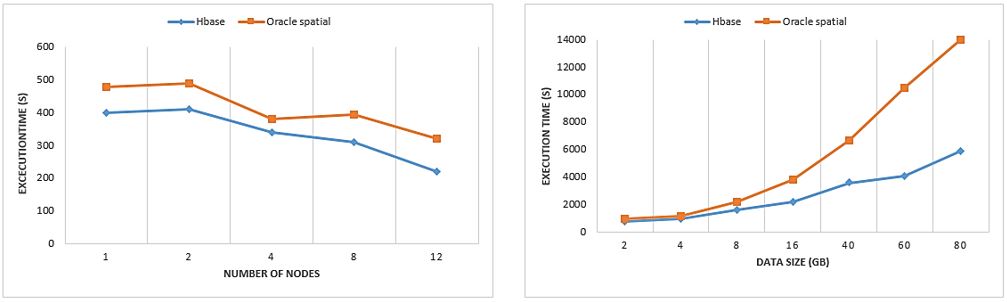}}
\caption{a.Data ingestion time of different worker nodes, b: Write time for different data size }
\label{Fig.8}
\end{figure}

Second, to measure the data size effect, the performance ingestion for each group of nodes is tested while increasing the dataset. 7 groups of images are used with different volume sizes starting from 2GB up to 80GB for each of the computing nodes. The images are divided into blocks, then stored into 12 nodes respectively. From Figure \ref{Fig.8}.a, it is noticeable that the time needed for data importation is reduced as the number of worker nodes is increased in the cluster for both Hbase and Oracle Spatial databases. The two databases scale well, yet the storage performance of the proposed model with Hbase is relatively more stable. According to this result, we conclude that as much as worker node instances are added in the cluster, better performance can be achieved, especially when there is a good network connection between the datanodes. From the other side, it can be observed from Figure \ref{Fig.8}.b that the time requested to insert the RS data image shows a small variance between two kinds of databases at the beginning, then increases relatively with the growth of RS image data size.
The performance of data importation of the proposed model is fairly more stable, and the difference between both databases is noticeable. Therefore, according to the experimental results the proposed tile-based distributed management model of massive high-resolution satellite images has been improved.

\section{Conclusion}
In this work, a distributed method for massive high-resolution image data storage and management is proposed based on the standard metadata file, the pyramid model, the Hilbert curve, and NoSQL databases management systems. Original images are divided into small blocks, then stored in Hbase, which is distributed in a group of nodes on the Hadoop framework. Experimental results prove the efficiency of the proposed model compared with traditional approaches. However, this model needs to be applied to other important applications such as image data retrieval and classification taking into account temporal-series RS image. In future works, we plan to use the proposed model for RS and IoT data ingestion. We also aim to develop a distributed DL algorithm for high-resolution image classification based on the proposed model for smart environment applications.

\bibliography{references}

\end{document}